# Work extraction, Information-content and the Landauer bound in the Continuous Maxwell Demon

M. Ribezzi-Crivellari[a,b] and F. Ritort[a,c]

[a]Condensed Matter Physics Department, University of Barcelona, C/Marti i Franques s/n, 08028 Barcelona, Spain.

[b]Laboratoire de Biochimie, Institute of Chemistry, Biology and Innovation (CBI), UMR 8231, ESPCI Paris/CNRS, PSL Research University, 10 rue Vauquelin, 75231-Paris Cedex 05, France

[c]CIBER-BBN de Bioingeniería, Biomateriales y Nanomedicina, Instituto de Sanidad Carlos III, Madrid, Spain.

**In a recent paper we have introduced a continuous version of the Maxwell demon (CMD) that is capable of extracting large amounts of work per cycle by repeated measurements of the state of the system [1]. Here we underline its main features such as the role played by the Landauer limit in the average extracted work, the continuous character of the measurement process and the differences between our continuous Maxwell demon and an autonomous Maxwell demon. We demonstrate the reversal of Landauer's inequality depending on the thermodynamical and mechanical stability of the work extracting substance. We also emphasize the robustness of the Shannon definition of the information-content of the stored sequences in the limit where work extraction is maximal and fueled by the large information-content of rare events.**

## 1. Introduction.

Despite its overwhelming presence in our everyday life information is among the less tangible physical quantities. Introduced by C. E. Shannon in 1948 information theory gave rise to a revolutionary area in modern science with implications in the most diverse fields, from communication theory in mathematics to quantum computation in physics and genetics in biology. Information theory also lies at the core of long debated questions such as the black hole information paradox in quantum mechanics [2] or the Maxwell demon paradox in statistical mechanics [3,4]. The latter paradox is considered being solved after the seminal works half a century ago in the framework of the thermodynamics of data processing by Landauer and Bennet [5-7]. During the recent years, and with the new possibilities offered by the invention of optical trapping and single molecule manipulation techniques [8,9], thermodynamics of information has spurred lots of research in the field of fluctuation theorems and information feedback [10-13].

The simplest model realization of a Maxwell demon (hereafter referred to as MD) is a Szilard engine [14]. This is a device that operates in contact with a thermal bath that is capable of extracting heat from the bath to fully convert it into work. In the classical Szilard engine (Figure 1a) a single particle occupies a vessel of volume $V$ made of two compartments $V_0, V_1$ $(V = V_0 + V_1)$. A measurement is made and the compartment occupied by the particle is determined. A wall is then inserted between the compartments and a work-extracting mechanism (e.g. a pulley) implemented to fully convert environmental heat into work. Let $P_0 = V_0/V$ and $P_1 = V_1/V$ $(P_0 + P_1 = 1)$ the probabilities to observe the particle in each compartment. The maximum extractable work is then $W_0 = -k_B T \log P_0$ and $W_1 = -k_B T \log P_1$ for compartments 0 and 1 and the maximum average work per measurement cycle equals,

$$W_{MD} = P_0 W_0 + P_1 W_1 = -k_B T (P_0 \log P_0 + P_1 \log P_1) \qquad (1)$$

which satisfies the Landauer limit $W_{MD} \leq W_L = k_B T \log 2$. Various experiments have validated the Landauer limit [15-20]. The standard Szilard corresponds to the classical version of the MD in which a given observation immediately leads to a work extraction process. However this is not the only way to realize an extractable work machine. One could think of alternative protocols where one makes repeated measurements and takes the decision to extract work only when a specific condition is met. In this case the average extractable work Eq.(1) no longer holds and other expressions are generally valid.

The experimental test of Eq.(1) in the classical Szilard engine requires implementing a *pulley* mechanism that maximizes the average extracted work. This is done by instantaneously changing a control parameter that stabilizes the state observed in the measurement, followed by an adiabatic recovery of the initial condition. In the classical Szilard [14] this is done by reversibly expanding the gas from the initial to the final volume against an opposing force. In the case of a bead in a double-well optical trap experiment [16] the stabilization of the observed state is done by unbalancing the double well potential by instantaneously lowering the free energy of the well occupied by the bead. In the case of the single electron box [18] the gate voltage is changed to stabilize the number of excess electrons in the two metallic islands. In the single DNA hairpin experiment reported in [1], the stabilization of the folded and unfolded state is produced by instantaneously moving the optical trap to instantaneously lower or increase the force applied on the molecular construct stabilizing the corresponding state (Figure 1b).

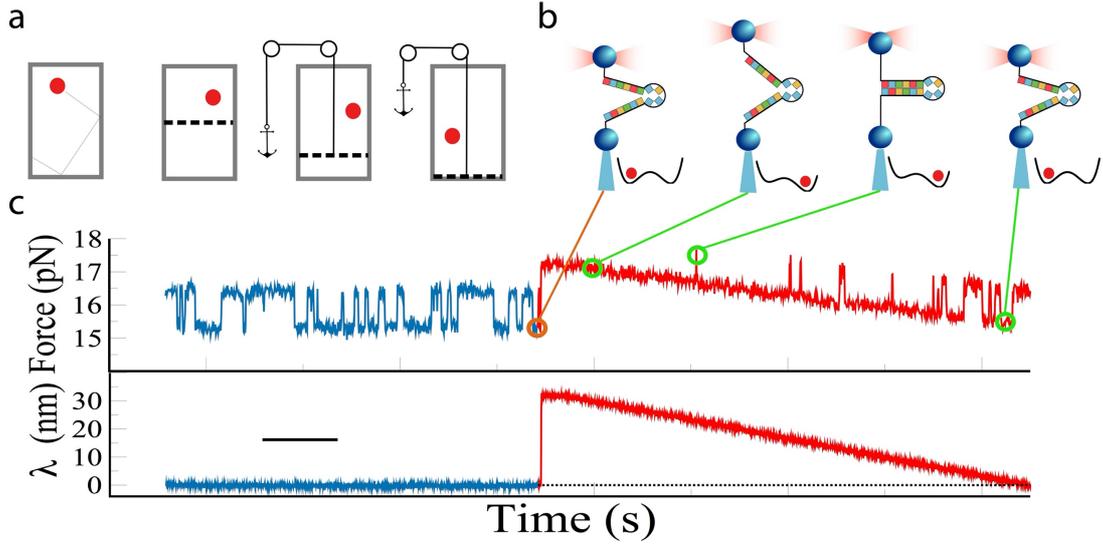

*Figure 1. (a) The Szilard engine as a work extracting machine. (b) Stabilization of the measured state along the work extraction cycle in a single DNA haiprin hopping between two states (folded and unfolded).*

## 2. The Continuous Maxwell demon (CMD) in a nutshell

In a recent paper [1] we have introduced the continuum Maxwell Demon (CMD) as a new conceptual framework to analyze the thermodynamics of data processing. At difference with the classical MD, the CMD monitors the time evolution of the system by performing repeated measurements of the system every time $\tau$ until the measurement outcome fulfills a given physical condition. Then a work-extracting machine is operated reversibly to fully convert heat into work. Results in the CMD were experimentally tested in a novel single-molecule Szilard motor DNA assay operating at room temperature. Four were the main results for the CMD in [1].

1. **The Landauer bound.** In the CMD the Landauer limit becomes a lower bound rather than an upper bound. The average work per measurement cycle is determined by the first measurement outcome. If it is 0 (1) then the extracted work equals $W_1(W_0)$ yielding a maximum average work per cycle,

$$W_{CMD} = P_0 W_1 + P_1 W_0 = -k_B T(P_0 \log P_1 + P_1 \log P_0) \qquad (2)$$

One can then readily demonstrate that $W_{CMD} \geq W_L = k_B T \log 2 \geq W_{MD}$, the Landauer limit becoming a lower bound in the CMD (rather than an upper bound). Note that Eqs.(1,2) have the same mathematical form, with just exchanged weights $P_0, P_1$. It is important to stress that the inversion of the Landauer inequality in the CMD is just a mathematical fact related to such exchange of the weights. The *new* inequality $W_{CMD} \geq W_L$ in the CMD complies with the Landauer principle and the Second Law as we show next.

2. **The Second Law.** The CMD can extract arbitrarily large amounts of work without violating the second law. In fact, Eq.(2) diverges in the limits $P_0 \to 0,1$. However the information-content of the stored sequences $I$ also does in that limit. In the classical MD a single bit is stored per measurement cycle, therefore $I_{MD} = -P_0 \log P_0 - P_1 \log P_1$ and $W_{MD} = k_B T I_{MD}$. In the CMD one can demonstrate $W_{CMD} \leq k_B T I_{min}$ with

$$I_{min} = -\frac{P_0}{P_1}\log(P_0) - \frac{P_1}{P_0}\log(P_1) - P_0 \log P_1 - P_1 \log P_0 \qquad (3)$$

being the minimum information in the CMD under lossless compression for repeated uncorrelated measurements [1]. These results agree with Landauer's principle and the second law. In the standard literature, the Landauer bound is formulated as $W \leq k_B T I$ with $I$ a proper information content, let it be Shannon information, mutual information or something like that, depending on the setup.

3. **Maximum efficiency.** The maximum efficiency $\epsilon_{max}$ is defined as the ratio between the maximum extracted work and the minimum energy needed to irreversibly erase the stored sequences,

$$\epsilon_{max} = \frac{W_{max}}{I_{min}} \qquad (4)$$

In the classical MD, $W_{MD} = k_B T I_{MD}$, and $\epsilon_{max} = 1$. Maximum efficiency in the CMD is found in the limit $P_0 \to 0,1$ where dynamics is ruled by rare events. In this limit, $W_{CMD} \to k_B T I_{min}$ both quantities diverging logarithmically like $-\log P_0$ (or $-\log(1 - P_0)$).

4. **Average power.** The CMD extracts large amounts of work per cycle as compared to the classical MD at the price of repeatedly measuring the state of the system until the particle changes compartment. Therefore the average duration of a cycle is always larger in the CMD as compared to the classical case, the latter being equal to $\tau$ (the time between consecutive measurements). In the classical MD the average power $P_{MD} = W_{MD}/\tau$, whereas in the CMD the average power is always lower, $P_{CMD} < P_{MD}$. Only in the limit $P_0 \to 0,1$ the average power in the CMD and the classical MD are asymptotically equal, both vanishing like $-P_0 \log P_0$ for $P_0 \to 0$. This result holds only at the level of the mean extracted power but does not hold at the level of probability distributions of the extracted work (or power) in a finite time [21]. The fact that two distributions have the same mean does not imply that the two distributions are identical.

5. **Experimental test.** The CMD has been experimentally tested in single DNA hairpin pulling experiments. In these experiments the molecule passively hops between the folded and an unfolded states until an observation is made (Figure 1b). Then a *pulley* mechanism is implemented whereby the optical trap is instantaneously moved to stabilize the observed state: if the molecule is observed in the folded state then the force is suddenly reduced; if the molecules is observed to be in the unfolded state then the molecular construct if further pulled and the force increased (Figure. 1c). In equilibrium conditions the probability of the two states 0 (folded) and 1 (unfolded) are given by, $P_0 = \frac{1}{1+e^\phi}$ and $P_1 = \frac{1}{1+e^{-\phi}}$ with $\phi$ being the equilibrium free energy difference (in $k_B T$ units) between the folded and the unfolded states: $\phi = (G_0 - G_1)/k_B T = \Delta G/k_B T = -\log\left(\frac{P_0}{P_1}\right)$. The average maximum work that can be extracted from each state is again given by $W_0 = -k_B T \log P_0$ and $W_1 = -k_B T \log P_1$ as in the Szilard gas.

The classical MD is the paradigm example of thermodynamics of information in discrete-time feedback (where the demon performs measurements at a given time). Although protocols with repeated measurement have been considered in the field of thermodynamics of information, they have been seldom implemented experimentally. In this regard the two-state model in [1] remains for now the simplest example where experiments and theoretical calculations for the information-content (c.f. Eq.(3)) can be worked out in detail.

Despite the strong activity in the field of thermodynamic information and the MD during the last decades, it is remarkable that no previous study has identified the simple implementation of a CMD as it has been done in [1]. The discovery of the CMD came during the course of a first series of information-to-energy conversion assays in DNA hairpins and after realizing that the standard version of the Maxwell demon admits a variant of the standard work extraction protocol that is equally conceptually challenging and, at the same time, experimentally accessible. This is just one example of how experiments spur further conceptual thinking and analytical work.

**3. The Landauer limit as a lower bound.**

The idea that the Landauer limit in the CMD becomes a lower bound (rather than an upper bound) has some interesting consequences. One may wonder how general this result is and whether it has implications on the thermodynamic and mechanical stability of the substance used to implement the Szilard engine (the gas molecule enclosed in the vessel for the Szilard model or the DNA hairpin hopping between the folded and unfolded states in the force unzipping experiments). These systems are both thermodynamically stable simply because they are in thermodynamic equilibrium. Are there stability conditions for the validity of the inequality $W_{CMD} > W_{MD}$? For instance, does it hold for substances that are thermodynamically unstable? Below we determine the conditions under which $W_{CMD} \geq W_{MD}$. In the classical MD the average work per cycle Eq.(1) is given by,

$$W_{MD} = P_0 W_0 + P_1 W_1 \leq k_B T \log 2 \qquad (5)$$

which is always lower than the Landauer limit $W_L = k_B T \log 2$. Instead, in the CMD the Landauer limit becomes a mathematical lower bound,

$$W_{CMD} = P_1 W_0 + P_0 W_1 \geq k_B T \log 2 \qquad (6)$$

Mathematically the only difference between Eqs.(5) and (6) is the exchange of weights, $P_0 \leftrightarrow P_1$ multiplying the quantities, $W_0, W_1$. Because $P_1 = 1 - P_0$, and $W_0$ and $W_1$ are functions of $P_0$ and $P_1$ respectively, $W_{MD}$ and $W_{CMD}$ in Eqs.(5,6) only depend on $P_0$. Let us generically denote $P_0$ by $P$. We claim that, among all physically acceptable mathematical functions $W(P)$, only a monotonically decreasing function of $P$ (such as the logarithmic form, $W \equiv -\log(P)$) simultaneously satisfies inequalities in Eqs. (5,6). The proof is as follows. Equations (5,6) can be rewritten as,

$$W_{MD}(P) = PW(P) + (1-P)W(1-P) \qquad (7)$$

$$W_{CMD}(P) = (1-P)W(P) + PW(1-P) \qquad (8)$$

with $W_{MD}(P = 1/2) = W_{CMD}(P = 1/2) = W(P = 1/2) = k_B T \log 2$ . The inequality $W_{CMD}(P) \geq W_{MD}(P)$ implies: $W(P) \geq W(1-P)$ if $P \leq 1/2$ and $W(P) \leq W(1-P)$ if $P \geq 1/2$. Since $0 \leq P \leq 1$ $W(P)$ must be a monotonically decreasing function of $P$ or $W'(P) < 0$ (we exclude here the trivial case $W'(P) = 0$ or $W(P) = constant$).

From a thermodynamic point of view the condition $W'(P) < 0$ implies that the system is mechanically stable, meaning that the pressure p for the Szilard gas or the force acting on the DNA hairpin is positive (i.e. pulling rather than pushing). In fact, a positive pressure $p = -\left(\frac{\partial G}{\partial V}\right)_T > 0$ means that the amount of the extracted work during a isothermal expansion of the Szilard gas, $V_i \rightarrow V_f (> V_i)$, is positive and given by $W = \int_{V_i}^{V_f} p dV = -\Delta G > 0$. A positive pressure p then gives $\Delta G < 0$, i.e. a positive work extraction is accompanied by a decrease in the free energy of the system.

In general, the probability $P$ of finding the molecule in a given compartment is monotonically increasing with its volume. For homogeneous substances one can assume $P \propto V$. Therefore $P$ is proportional to the initial volume $V_i$ in the isothermal expansion of the work extraction process, $\frac{\partial P}{\partial V_i} = const > 0$. From $W = \int_{V_i}^{V_f} p\, dV = -\Delta G > 0$ we get

$$W'(P) = \frac{\frac{\partial W}{\partial V_i}}{\frac{\partial P}{\partial V_i}} = -\frac{p}{\frac{\partial P}{\partial V_i}} < 0 \qquad (9)$$

Equation (9) is the *mechanical stability condition*. Let us note that mechanical stability is not necessarily accompanied by *thermodynamic stability*, the latter implying that $W(P)$ is a convex function, $W''(P) > 0$. In fact, thermodynamic stability implies that the isothermal compressibility is positive, $\kappa_T = \frac{1}{\left[V\left(\partial^2 G/\partial V^2\right)_T\right]} > 0$. Using again $W = -\Delta G$ and $\frac{\partial P}{\partial V_i} = const > 0$ we get $W''(P) > 0$. Therefore $W_{CMD} \geq W_{MD}$ still holds for substances that are mechanically stable but thermodynamically unstable ($W''(P) < 0$).

In Figure 2 we show $W_{CMD}(P)$ and $W_{MD}(P)$ for three examples all mechanically stable ($W'(P) < 0$) but one thermodynamically unstable $W''(P) < 0$.

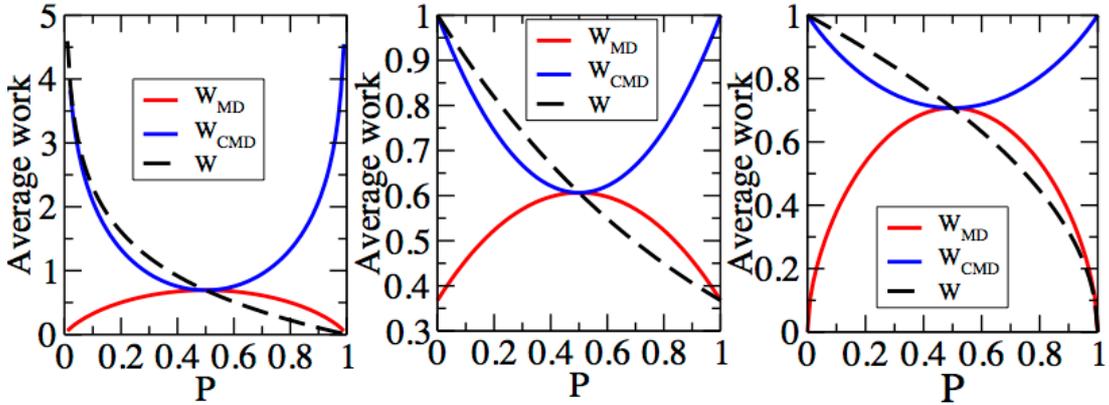

**Figure 2.** *Functions $W(P)$, $W_{MD}(P)$ and $W_{CMD}(P)$ for three cases with mechanical stability, $W'(P) < 0$. (Left) $W(P) = -\log P$, the case of the Szilard model considered in [1]. (Middle) $W(P) = \exp(-P)$. (Right) $W(P) = \sqrt{1-P}$. Left and middle are mechanical and thermodynamic stable. Right is mechanically stable but thermodynamically unstable. In all three cases we have $W_{CMD} > W_{MD}$.*

Summing up, the mechanical stability condition $W'(P) < 0$ implies $W_{CMD} > W_{MD}$. Conversely, if the working substance is mechanically unstable ($W'(P) < 0$) then $W_{CMD} \leq W_{MD}$ and the Landauer limit is an upper bound for both the classical MD and the CMD.

## 4. Information-content of the stored sequences in the CMD

The calculation of the information-content of the stored sequences for the CMD presented in [1] follows standard Shannon information theory. To compute this one considers a sum over all possible multiple bit sequences corresponding to a given cycle. Let $\Gamma$ denote a path or trajectory made of *n+1* measurements corresponding to a work extractable cycle of the class $\mathbf{0}_n = \{\overbrace{0,\ldots 0}^{n}, 1\}$ or $\mathbf{1}_n = \{\overbrace{1,\ldots 1}^{n}, 0\}$, i.e. $\Gamma \in \{\mathbf{0}_n, \mathbf{1}_n\}$ with $n \geq 1$. Sequences $\mathbf{0}_n, \mathbf{1}_n$ are such that the first $n$ measurements yield the same outcome followed by the last (*n+1*) different outcome. Work distributions for the work extraction process are given by,

$$P(W) = \sum_{\Gamma} P(\Gamma)\, \delta(W - W(\Gamma))$$

$$= \sum_{\Gamma \in \{\mathbf{0}_n\}} P(\Gamma)\, \delta(W - W_1) + \sum_{\Gamma \in \{\mathbf{1}_n\}} P(\Gamma)\, \delta(W - W_0) =$$

$$P_0 Q_1(W) + P_1 Q_0(W) \qquad (10)$$

where $Q_0(W), Q_1(W)$ are the extracted-work distributions conditioned to the system being observed in states 0,1 and $W_0, W_1$ are the mean. This yields Eq.(2), $W_{CMD} = P_0 W_1 + P_1 W_0$. It is important to note that $P(W)$ in Eq.(10) and the values of $W_0, W_1$ are independent of the time $\tau$ between consecutive measurements.

The average information-content we calculated is given by,

$$I = -\sum_{\Gamma} P(\Gamma) \log P(\Gamma) \qquad (11)$$

This is the equivalent of the statistical entropy but calculated over paths of stored bits. These quantities are standard in the field of statistical mechanics and have being used in different contexts [22-24]. As shown in [1], an in contrast to the work distribution (9), $I$ depends on the time $\tau$ between consecutive measurements. For $\tau$ much larger than the decorrelation time of the system, the information-content converges to $I_{min}$ (Eq.(3))

In order to substantiate the connection between Eq.(11) and Shannon entropy we present an alternative derivation of the main result Eq.(3) based on the optimal code length of a sequence of bits [25]. Albeit approximate it works extremely well, becoming exact in the limit $P_0 \to 0,1$. Since the work by Landauer and others we know that the work output of a Maxwell's Demon must be balanced by the cost of erasing the information acquired during the Demon's action. In our case the protocol requires repeated (N+1) measurements. However, by construction, the first N of these N+1 measurements will yield the same result (Figure 3, top left). This means that the string obtained from our measurement can be encoded without loss using one bit for the first measurement (defining the cycle) and recording the total length (N) of the measurement. Lossless re-encoding is a logically reversible operation and, as such, can be performed reversibly [6].

Writing N in binary requires $N_{comp}=\lfloor \log_2(N) + 1 \rfloor$ bits, where $\lfloor ... \rfloor$ denotes the integral part (the so-called floor of a real number) and $\log_2$ is the logarithm in base 2 (Figure 3, bottom left). The average of this quantity (defined by $<\cdots>$) over different realizations (i.e. different sequences) cannot be computed analytically but can be computed numerically by generating a large number of dynamical sequences to arbitrary precision. According to our argument the average information-content of the stored sequences is then $I = <N_{comp} + 1> \log 2 = <N_{bit}> \log 2$. This quantity is compared to $I_{min}$ (Eq.3) in Figure 3 (bottom). There we show that the $<N_{bit}> \log 2$ is a good approximation to $I_{min}$, the two expressions differing by less than one bit over the range of possible values of $P_0$ ($I_{min}$ being bounded between $<N_{bit}> \log 2$ and $<N_{bit} + 1> \log 2$). Thus the Shannon entropy, $I_{min}$, has a clear though approximate interpretation in terms of the number of bits required to store the measurement result under lossless compression. The relative error of this approximation is less than 10% on the whole range of values of $P_0$ vanishing in the relevant limit $|\phi| \gg 1$ where dynamics is dominated by rare events.

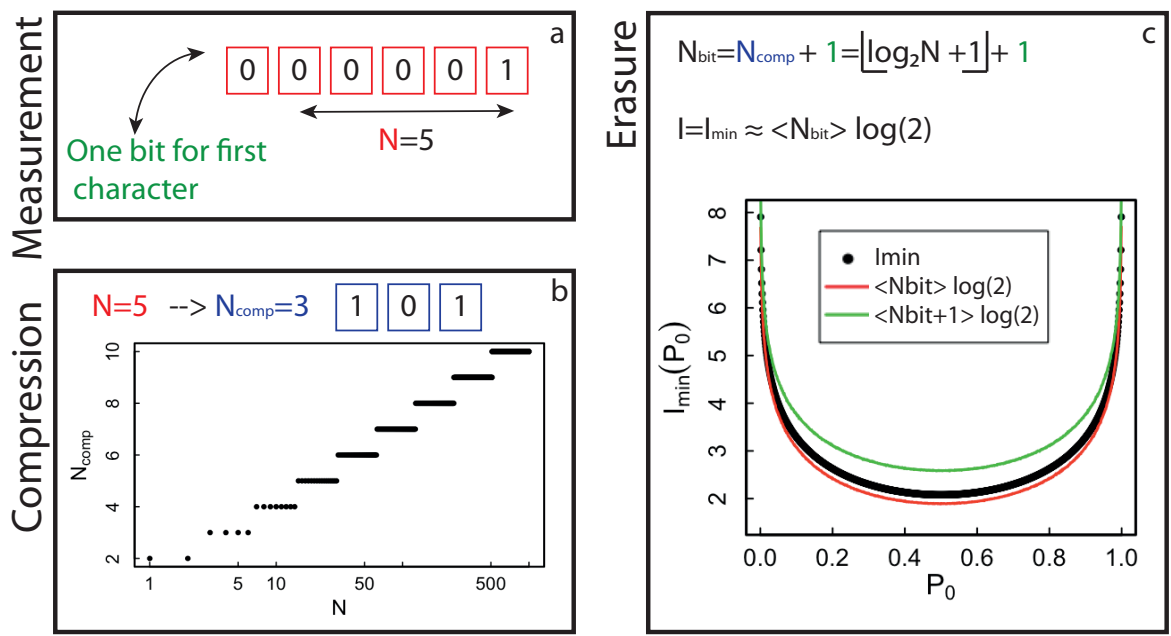

**Figure 3.** *An approximated solution for $I_{min}$ based on optimal coding of a sequence of bits.*

## 5. The Continuous Maxwell Demon versus the Autonomous Maxwell demon

In the CMD the demon makes repeated measurements of the state of the system every finite time $\tau$ until a physical condition is met: in our case the system (particle in a gas or molecular conformation) changes state. The qualification of *continuous* Maxwell Demon stands to indicate that $\tau$ can be arbitrarily small. This is in stark contrast to the classical MD where a single measurement is made and there is no characteristic measurement timescale $\tau$. Truly speaking the continuous case in the CMD should refer to the limit $\tau \to 0$, however analytical calculations in a Markov process and experimental measurements in the CMD make sense only for finite $\tau$, the continuous-time limit being understood in the limit $\tau \to 0$. From a practical point of view measurements can be taken repeatedly at any achievable rate (given by the experimental limitations). However the limit $\tau \to 0$ turns out to be unphysical because in that limit the number of bits diverges as $1/\tau$ and no physical memory is capable of storing an infinite amount of information. Indeed after lossless compression the information-content of correlated bit sequences diverges logarithmically as $-\log(R\tau)$ [1]. On the other hand, according to Landauer, the energy needed to erase an infinite amount of stored information would be also infinite. Therefore, any CMD must operate by measuring at potentially highly frequent but repeated discrete times.

The CMD should be distinguished from the so-called autonomous Maxwell Demon (AMD). The AMD involves two or more physically coupled systems rather than just a single system and a work-extracting machine as in the Szilard engine [26]. The general scheme of an AMD is as follows. Two physically coupled systems X and Y are kept in contact with a thermal bath. System Y interacts with system X and, when X changes conformation, Y reacts to that change. Now, let us say that system Y (the demon) is designed to react in a specific way upon the state of X (the system). If system X is driven out of equilibrium, the continued interaction between X and the *reactive* demon Y will modulate the nonequilibrium steady reached by X. Remarkably this can be used to cool X to a lower temperature. The steady state reached by X is a direct consequence of the information continuously exchanged between systems X and Y. In this regard the AMD might be considered as a truly *continuous-time* machine as the physical interaction between systems X and Y continuously proceeds in time. The proper information-related quantity in the AMD have been suggested to be mutual information and the transfer entropy [27]. Transfer entropy quantifies the mutual-information rate change between two physically coupled systems, X and Y, after one additional measurement is made in either X or Y. Mutual information measurements in an AMD have been implemented in single electron devices [28-30] by coupling two single electron boxes. In this case one box (the demon) is designed to react to the state of the other box (the system). No work extraction process is implemented on the system and the reaction protocol is designed in such a way that the system *cools* down while the demon heats up, while both system and demon remain in contact with a thermal bath.

## 5. Concluding remarks

Half a century ago it was shown that any irreversible logical operation, such as bit erasure, requires energy consumption typically on the order of k$_B$T [5,6]. This simple fact resolves the apparent violation of the second law in the Maxwell demon (MD) and Szilard engine paradoxes. The experimental realization of the Maxwell demon or the Szilard engine has seen an upsurge of interest thanks to the development of technologies capable of manipulating small systems such as colloidal particles, single molecules or miniaturized electronic devices. In the classical MD operating as a Szilard engine the work extraction process is implemented at every observation and a single bit is required per work extraction cycle. In the continuous MD (CMD) observations are repeatedly made until a condition is met (e.g. the particle changes compartment or the molecule changes conformation) and multiple bits must be stored per work extraction cycle. In the classical MD the Landauer limit ($k_B T \log 2$) equals the information-content of a one-bit measurement. In contrast, the reversal of the Landauer inequality in the CMD is consequence of the fact that extracting work for the less probable state is always advantageous with respect to extracting it from the more probable state. In turn, this is possible because of the larger information-content of the multiple bits stored sequences in the CMD, thus saving the second law, $W_{CMD} \leq k_B T I_{min}$. Here we have shown that the result $W_{CMD} \geq k_B T \log 2$ is always true whenever the working-extracting substance is in mechanical equilibrium. Thermodynamic equilibrium being not a necessary condition renders this result even more general. The CMD can extract arbitrarily large amounts of work (at the price of an equally large information storage) beyond the Landauer limit and providing a new conceptual framework to think about information in physics. The CMD might be particularly relevant for regulatory processes in biology that operate under similar conditions (e.g. when the concentration of a chemical or the voltage across a synapsis reaches a threshold value, see e.g. [31]).

A legitimate question is whether the information-content calculation in Eq.3 [1] is unique and whether one could exactly match the average extracted work Eq.(2). Matching $W_{CMD}$ in Eq.(2) with $I_{min}$ in Eq.(3) would result in a maximally efficient CMD with $\epsilon_{max} = 1$ (c.f. Eq.(4)) as in the classical MD. In principle, one might encode repeated measurements in a more clever way than just annotating bits in a sequential fashion. Mathematically Eqs.(2,3) only differ in the term $-\frac{P_0}{P_1}\log(P_0) - \frac{P_1}{P_0}\log(P_1)$ present for $I_{min}$ in Eq.(3). However it is unclear whether there is a way to encode the multiple observations (for finite $\tau$ or in the limit $\tau \to \infty$) such that the Shannon entropy of the stored sequences equals the maximum average work Eq.(2). This remains an interesting open question. The same issue arises in many other examples of multiple measurements under repeated feedback control [32-35]

One of the most relevant results in [1] is that $I_{min}$ in Eq.(3), and more in general the information-content of the stored sequences for an arbitrary finite measurement time $\tau$, saturates the average maximum extractable work (i.e. the efficiency Eq.4 goes to 1) in the limit $|\phi| = \log\left(\frac{P_0}{P_1}\right) \gg 1$ (or $P_0 \to 0, 1$). Several information bounds have been proposed for multiple repeated measurements in statistical systems [36] raising questions about their interpretation and utility. A key result in all our calculations, exact or approximate (e.g. the one presented in Section 4 based on estimating the optical code length of a sequence of bits), is the appearance of the non-trivial and divergent term the limit $|\phi| \gg 1$ (or $P_0 \to 0, 1$): $-P_0 \log P_1 - P_1 \log P_0$ in the expression of the information-content (Eq.3). Any entropy definition (transfer entropy, information flow, mutual information among others, see [36] for details) incapable of yielding such mathematical term will never attain the efficiency of 1 so characteristic of the CMD in the limit $|\phi| \gg 1$.

**Acknowledgements.** We acknowledge financial support from Grants 308850 INFERNOS, 267862 MAGREPS (FP7 EU program) FIS2013-47796-P, FIS2016-80458-P (Spanish Research Council) and Icrea Academia prize 2018 (Catalan Government). M.R.-C. has received funding from the EU Horizon 2020 research and innovation programme under the Marie Sklodowska-Curie grant agreement No. 749944.

# References

[1] M. Ribezzi-Crivellari and F. Ritort, *Large work extraction and the Landauer limit in a continuous Maxwell demon*, Nature Physics, Doi: 10.1038/s41567-019-0481-0 (2019)

[2] S. B. Giddings, *Black holes, quantum information, and the foundations of physics,* Phys. Today **66,** 30 (2013)

[3] H. S. Leff and A. F. Rex, ed. *Maxwell's Demon: Entropy, Information, Computing*. Bristol: Adam-Hilger (1990)

[4] E. Lutz and S. Ciliberto, *Information: From Maxwell's demon to Landauer's eraser,* Phys. Today **68,** 30 (2015)

[5] R. Landauer, *Irreversibility and heat generation in the computing process*, IBM J. Res. Develop. **5**, 183-191 (1961)

[6] C. H. Bennett, *The thermodynamics of computation: a review*, Int. J. Theor. Phys. **21**, 905-940 (1983)

[7] C. H. Bennet and R. Landauer, *The fundamental Physical limits of computation*. Scientific American **253,** 48 (1985)

[8] F. Ritort, *Single molecule experiments in biological physics: methods and applications*, Journal of Physics C (Condensed Matter) **18,** R531-R583 (2006).

[9] S. Ciliberto, *Experiments in Stochastic Thermodynamics: Short History and Perspectives*, Physical Review X **7**, 021051 (2017)
[10] T. Sagawa and M. Ueda, *Generalized Jarzynski equality under nonequilibrium feedback control*, Phys. Rev. Lett. **104**, 090602 (2010)
[11] U. Seifert, *Stochastic thermodynamics, fluctuation theorems, and molecular machines*, Rep. Prog. Phys. **75**, 126001 (2012)
[12] T. Sagawa, *Thermodynamic and logical reversibilities revisited*, J. Stat. Mech. P03025 (2014)
[13] J. M. R. Parrondo, J. M. Horowitz and T. Sagawa, *Thermodynamics of information,* Nat. Phys. **11**, 131 (2015)
[14] L. Szilard, *On the decrease of entropy in a thermodynamic system by the intervention of intelligent beings,* Z. Phys. **53**, 840 (1929)
[15] S. Toyabe, T. Sagawa, M. Ueda, E. Muneyuki and M. Sano, *Experimental demonstration of information-to-energy conversion and validation of the generalized Jarzynski equality,* Nat. Phys. **6**, 988 (2010)
[16] A. Berut, A. Arakelyan, A. Petrosyan, S. Ciliberto, R. Dillenschneider and E. Lutz, *Experimental verification of Landauer's principle linking information and thermodynamics,* Nature **483,** 187-190 (2012)
[17] Y. Jun, M. Gavrilov and J. Bechhoefer, *High-Precision Test of Landauer's Principle in a Feedback Trap,* Phys. Rev. Lett. **113**, 190601 (2014)
[18] J. V. Koski, V. F. Maisi, J. P. Pekola and D. V. Averin, *Experimental realization of a Szilard engine with a single electron,* Proc. Natl. Acad. Sci. (USA) **111,** 13786-13789 (2014)
[19] J. Hong, B. Lambson, S. Dhuey, J. Bokor, *Experimental test of Landauer's principle in single-bit operations on nanomagnetic memory bits,* Sci. Adv. **2**, e1501492 (2016)
[20] J. P. S. Peterson, R. S. Sarthour, A. M. Souza, I. S. Oliveira, J. Goold, K. Modi, D. O. Soares-Pinto and L. C. Celeri, *Experimental demonstration of information to energy conversion in a quantum system at the Landauer limit,* Proc. R. Soc. A **472**, 20150813 (2016)
[21] G. Verley, M. Esposito, T. Willaert and C. Van den Broeck, *The unlikely Carnot efficiency,* Nat. Commun. **5**, 4721 (2014)
[22] F. Ritort, *Work and heat fluctuations in two-state systems: a trajectory thermodynamics formalism,* JSTAT P10016 (2004)
[23] A. Imparato and L. Peliti, *Work distribution and path integrals in mean-field systems*, Europhys. Lett. **70,** 740–6 (2005)
[24] M Merolle, J. P. Garrahan and D. Chandler, *Space–time thermodynamics of the glass transition*, Proceedings of the National Academy of Sciences **102,** 10837-10840 (2005).
[25] T. M. Cover and J. A. Thomas, *Elements of Information Theory,* John Wiley & Sons, Inc. (1991)
[26] D. Mandal and C. Jarzynski, *Work and information processing in a solvable model of Maxwell's demon*, Proc. Natl. Acad. Sci. (USA) **109,** 11641-11645 (2012)
[27] T. Schreiber, *Measuring information transfer,* Phys. Rev. Lett. **85**, 461 (2000)
[28] P. Strasberg, G. Schaller, T. Brandes and M. Esposito, *Thermodynamics of a physical model implementing a Maxwell demon,* Phys. Rev. Lett. **110**, 040601 (2013)
[29] J. V. Koski, A. Kutvonen, I. M. Khaymovich, T. Ala-Nissila and J. P. Pekola, *On-Chip Maxwell's demon as an Information-Powered Refrigerator,* Phys. Rev. Lett. **115,** 260602 (2015)
[30] K. Chida, S. Desai, K. Nishiguchi and A. Fujiwara, *Power generator driven by Maxwell's Demon,* Nature Communications **8**, 15310 (2017)
[31] S. Ito and T. Sagawa, *Maxwell's demon in biochemical signal transduction with feedback loop,* Nature Communications **6**, 7498 (2015)
[32] J. M. Horowitz and S. Vaikuntanathan, *Nonequilibrium detailed fluctuation theorem for repeated discrete feedback,* Physical Review E **82** 061120 (2010)
[33] R. K. Schmitt, J. M. R. Parrondo, H. Linke and J. Johansson, *Molecular motor efficiency is maximized in the presence of both power-stroke and rectification through feedback,* New Journal of Physics 17, 065011 (2015)
[34] T. Admon, S. Rahav, and Y. Roichman, *Experimental Realization of an Information Machine with Tunable Temporal Correlations*, Physical Review Letters **121**, 180601 (2018)
[35] M. Debiosac, D. Grass, J. J. Alonso, E. Lutz and N. Kiesel, *Thermodynamics of continuous non-Markovian feedback control,* Preprint arXiv :1904.04889v1 [quant-ph]
[36] J. M. Horowitz and H. Sandberg, *Second-law-like inequalities with information and their interpretations,* New J. Phys. **16,** 125007 (2014)